\documentstyle[epsf]{article}
\epsfverbosetrue
 
\font\tenrm=cmr10
\font\tenit=cmti10
\font\elevenbf=cmbx10 scaled\magstep 1
\font\elevenrm=cmr10 scaled\magstep 1
 1

\def\npb#1#2#3{    {\it Nucl. Phys. }{\bf B\,#1} (19#2) #3}
\def\plb#1#2#3{    {\it Phys. Lett. }{\bf B\,#1} (19#2) #3}
\def\prd#1#2#3{    {\it Phys. Rev. }{\bf D\,#1} (19#2) #3}

\def\prl#1#2#3{    {\it Phys. Rev. Lett. }{\bf #1} (19#2) #3}

\def\zpc#1#2#3{    {\it Zeit. f\"ur Physik }{\bf C\,#1} (19#2) #3}

\def\ltap{\ \raisebox{-.4ex}{\rlap{$\sim$}} \raisebox{.4ex}{$<$}\ }

\def\beq{\begin{equation}}
\def\eeq{\end{equation}}
\def\bea{\begin{eqnarray}}
\def\eea{\end{eqnarray}}

\newcommand{\wti}{\widetilde}

\newcommand{\mch}{\mbox{$ m_{H^+}$}}
\newcommand{\tanb}{\mbox{$\tan \! \beta$}}
\newcommand{\bsg}{\mbox{$b \rightarrow s \gamma$}}

\newcommand{\thb}{\mbox{$t\to H^+ b$}}
\newcommand{\tstn}{\mbox{$t\to {\wti u}_1 {\wti \chi}^0_1$}}
\newcommand{\bctaunu}{\mbox{$b \to c \tau \nu$}}
\newcommand{\Htn}{\mbox{$H^+\to \tau^+ \nu_\tau$}}
\newcommand{\Hcn}{\mbox{$H^+\to {\wti \chi}_2^+ {\wti \chi}_1^0$}}
\newcommand{\gamw}{\mbox{$\Gamma(t \to W^+ b)$}}
\newcommand{\gamh}{\mbox{$\Gamma(t \to H^+ b)$}}

\begin{document}

\begin{titlepage}
\noindent
\phantom{a}     \hfill         WIS-96/13/Feb.-PH    \\
\phantom{a}     \hfill         TUM-T31-87/95        \\ 
\phantom{a}     \hfill         LMU-TPW-96-12        \\
\phantom{a}     \hfill         February 1996        \\[7ex]

\begin{center}

{\bf SUPERSYMMETRIC DECAYS OF THE TOP QUARK: AN UPDATE}
\footnote{Contribution to the  {\it $e^+e^-$ 2000\,GeV Linear 
   Collider Workshop}, Annecy, Gran Sasso, Hamburg, 
   February 1995--August 1995 }                            \\[11ex]
{\bf    Francesca M.\ Borzumati     }                      \\[1ex]
{\it Inst. Theoret. Physik, Techn. Universit\"at M\"unchen, 
     Garching, Germany                                   } \\
{\it Physics Department, Weizmann Institute, Rehovot, 
               Israel                                    } \\[5ex]
{\bf    Nir\ Polonsky     }                                \\[1ex]
{\it   Sekt. Physik Universit\"at M\"unchen,                       
    L.S. Prof.~Wess, M\"unchen, Germany                  } \\[12ex]

\end{center}
{\begin{center} ABSTRACT \end{center}}
\vspace*{1mm}

%\parbox{13.5cm}
{
\noindent
We analyze the two decays $\thb$, $\tstn$ within the Minimal
Supersymmetric Standard Model with radiative breaking of the
electroweak sector. We discuss their detectability at present and in
the eventuality that supersymmetry is not discovered at LEPII.
%\end{center}
}
\vfill
\end{titlepage}

\thispagestyle{empty}
\phantom{aa}
\newpage

\setcounter{page}{1}
\begin{center}
{\elevenbf SUPERSYMMETRIC DECAYS OF THE TOP QUARK: AN UPDATE} 
\vskip 0.6truecm
{\tenrm FRANCESCA M.~BORZUMATI$^{1,2}$, N.~POLONSKY$^3$ } 
\vskip 0.05truecm 
\baselineskip=13pt
{\tenit  $^1$
   Inst. Theoret. Physik, Techn. Universit\"at M\"unchen, 
    Garching, Germany                                      \\}
{\tenit  $^2$
   Physics Department, Weizmann Institute, Rehovot, Israel \\}
{\tenit  $^3$
   Sekt. Physik Universit\"at M\"unchen,                       
    L.S. Prof.~Wess, M\"unchen, Germany                      }
\vskip 0.4truecm 
{\tenrm ABSTRACT}
\end{center}
\vskip 0.1truecm
{\rightskip=3pc \leftskip=3pc \tenrm\baselineskip=10pt
We analyze the two decays $\thb$, $\tstn$ within the Minimal
Supersymmetric Standard Model with radiative breaking of the
electroweak sector. We discuss their detectability at present and in
the eventuality that supersymmetry is not discovered at LEPII.
}
\vskip 0.5truecm 
{\elevenbf\noindent 1. Problem and Inputs}
\vskip 0.2truecm 
\baselineskip =14pt 
\elevenrm 

\looseness=-1
It is well known that in supersymmetric models the top quark can decay
at the tree-level into a charged Higgs plus a bottom quark, $\thb$,
and into a stop ${\wti u}_1$ plus the lightest neutralino 
${\wti \chi}_1^0$, $\tstn$ (see for example~\cite{FBtop} and 
references therein). Both decays can have sizable rates: \thb\ can 
easily reach the $10\!-\!30\%$ level for large \tanb; $\tstn$ has 
similarly large rates in some corners of the supersymmetric 
parameter space~\cite{FBtop}.

\looseness=-1
The agreement between the top production cross section measuread at
the TEVATRON~\cite{CDFtop} and that which is predicted by the Standard
Model (SM) still allows such large rates. Moreover, the measurement of
the top mass ($M_t$) based on electron and muon tagging is not
sensitive to these decay modes. For very large values of \tanb\, which
give large \thb\ rates, $H^+$ decays into third generation leptons
$\tau \nu_\tau$ (more than $95\%$ of the times), whereas, in general,
the decay into the lightest chargino (${\wti \chi}_2^+$) and
neutralino, \Hcn\ , may dominate. In the second mode $\tstn$, 
${\wti u}_1$ can decay as ${\wti u}_1 \to {\wti \chi}_2^+ b$ and 
${\wti u}_1 \to c {\wti \chi_1}^0$. In all these channels, a larger 
amount of missing energy than in the standard $t \to W^+ b$ is
produced, which leads to softer spectra for the charged leptons.

\looseness=-1
By phase space suppression, the abovementioned rates can be 
progressively reduced to the level of ``rare'' ones when the masses 
of $H^+$, ${\wti u}_1$, and ${\wti \chi}_1^0$ increase. The 
question to be answered, therefore, is whether the Minimal
Supersymmetric Standard Model (MSSM) can still support these two decay
channels (and at which level), once all constraints coming from
experiments are imposed. Independently, one should also consider how
much information can be extracted for possible extensions of the MSSM
by the observation (or non-observation) of these decay modes. In this
contribution we provide an answer only to the first of these two
questions.

\looseness=-1
The MSSM studied here is the most minimal realization of a
supersymmetric version of the SM, with breaking of the electro-weak
sector radiatively induced. We assume that the universal boundary
conditions are given near the grand-unified scale of 
$M_{G} \sim 3 \cdot 10^{16}$ GeV. We do not make any assumptions 
regarding the details of the physics at the grand scale and we do 
not assume $b \!- \!\tau$ unification, which is affected by these
details~\cite{NPunif}. We fix the QCD coupling constant to its
average value $\alpha_{s}(M_{Z}) = 0.12$ and the running mass 
$m_b(M_Z)$ to be $m_{b}(M_{Z}) = 3\,$GeV.

\looseness=-1
We include radiative corrections to the Higgs potential following 
the procedure described in~\cite{NPsimul} while we impose 
radiative breaking of the electroweak sector. We find that the spectrum 
of the Higgs mass parameters calculated using one-loop 
renormalization group equations and tree-level sum rules is slightly 
heavier than that which is described in~\cite{FBtop,FBbsg} where these 
corrections were not included.

\looseness=-1
Furthermore, we evaluate the mass--shifts for $H^+$ induced by these
corrections as in~\cite{BB}. We compare these
results with those obtained following other calculations which make
use, as~\cite{BB}, of the diagrammatic approch~\cite{HH} and of the
effective potential method~\cite{BERZ,DRNO}. We find, in general, a
rather good agreement (after replacing $h_b$ with $-h_b$ in (5d) and
(5e) of~\cite{DRNO}). In generic supersymmetric models these mass
corrections can be large for large left-right mixings in the squark
mass matrices and can be positive and negative (this last possibility
is in general observed for small values of \tanb). In our MSSM
simulation, however, after radiative breaking of the electroweak sector
is imposed, we find that the corrected masses hardly deviate from
those obtained using the tree-level sum rules.

\looseness=-1
We come now to discuss the cuts imposed to the region of parameter
space which we study. It is known that the decay \bctaunu\ restricts 
the ratio $r=\tanb \,({\rm GeV})/\mch$ to be $\ltap 0.5$ in models 
containing 
two Higgs doublets with couplings of type~II to the 
fermions (see~\cite{GHNL} and related references therein). Although 
relevant for models of
global realizations of supersymmetry and/or extensions of the
MSSM~\cite{KaneWells}, this decay is completely ineffective in our 
case. As we shall see,
the experimental lower bounds on supersymmetric masses and the
constraints coming from the requirement of radiative breaking of the
electroweak gauge group push already \mch\ to be
larger than \tanb\ (in GeV). 

\looseness=-1
Similarly, the combined limit $\mch$--$\tanb$ recently set by 
CDF~\cite{CDFch} by measuring energetic jets coming from b-quarks 
and hadronically decaying $\tau$'s (only the mode \Htn\ is 
considered), does not 
affect the regions of parameter space which we obtain in our 
MSSM simulation. 

\looseness=-1
As in~\cite{FBtop}, also in this analysis, the only effective 
experimental bounds are those coming from direct searches of 
supersymmetric particles and from the decay \bsg.

\looseness=-1
In agreement with results coming from LEP~I, LEP1.5~\cite{GUSTAVO}, 
and the TEVATRON, the cuts which we apply to the masses of gluinos 
$\wti g$, charginos ${\wti \chi}^-$,
neutralinos ${\wti \chi}^0$, charged and neutral sleptons $\wti l$,
$\wti \nu$, up- and down-squarks $\wti u$,~$\wti d$, and neutral
Higgses (with $h_2^0$ the lightest of the two CP-even states, $h_3$
the CP-odd state and $H^\pm$ the charged Higgs) are (in GeV):
$ m_{\wti g}                \!>\!   150$,
$ m_{{\wti d_1},{\wti u_2}} \!>\!   150$,
$ m_{\wti u_1}              \!>\!    45$,
$ m_{\wti\chi^-_2}          \!>\!    60$,
$ m_{\wti\nu_1}             \!>\!    45$,
$ m_{\wti l_1}              \!>\!    45$,
$ m_{h_2^0}                 \!>\!    45$,
$ m_{h_3^0}                 \!>\!    20$,
$ m_{H^\pm}                 \!>\!    45$,
$ m_{\wti\chi^0_1}          \!>\!    20$.
For the limit on the mass of the stop-squark we rely on searches 
at LEP since the decay ${\wti u_1} \to t + {\wti \chi^0_1}$ is 
forbidden or disfavoured at the TEVATRON. The limit on the mass of
${\wti \chi^0_1}$, in general higher than the one obtained at
LEP, is naturally induced in the MSSM by the limit on the 
${{\wti \chi_2}^-}$--mass. Finally, the inclusion of 
radiative corrections to the Higgs potential and the subsequent 
modification of the tree-level sum rules for the Higgs-masses 
force us to impose individual lower bounds for all Higgs
particles.

\looseness=-1
The situation which the Next Linear $e^+e^-$ Collider may have to face
if supersymmetric particles remain undiscovered at LEP~II can be
mimicked by imposing the following bounds~\cite{LEPII} (in GeV):
$ m_{\wti g}                  \!>\! 200$,
$ m_{{\wti d_1},{\wti u_2}}   \!>\! 200$,
$ m_{\wti u_1}                \!>\!  85$,
$ m_{\wti\chi^-_2}            \!>\!  85$,
$ m_{\wti\nu_1}               \!>\!  85$,
$ m_{\wti l_1}                \!>\!  85$,
$ m_{h_2^0}                   \!>\!  85$,
$ m_{h_3^0}                   \!>\!  40$,
$ m_{H^\pm}                   \!>\!  85$,
$ m_{\wti\chi^0_1}            \!>\!  40$\,. 
We assume that the TEVATRON limits on squarks and gluino masses 
will not increase more than $50\,$GeV. 
In the following we refer to these two choices of 
bounds as SET~I and SET~II. 

\begin{figure}[t]
\epsfxsize=8.0 cm
\leavevmode
\epsfbox[50 360 550 745]{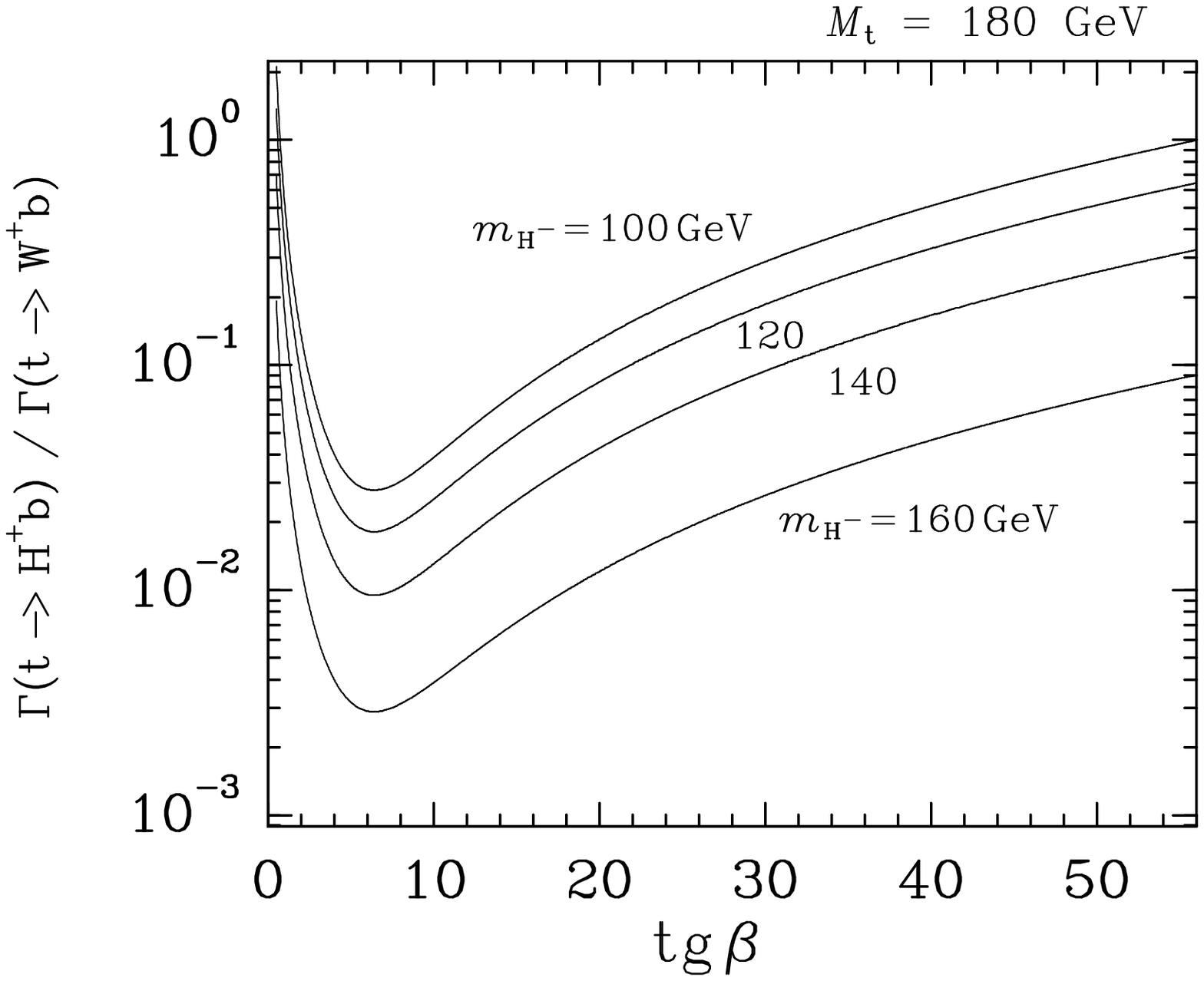}
\epsfxsize=8.0 cm
\epsfbox[30 360 530 745]{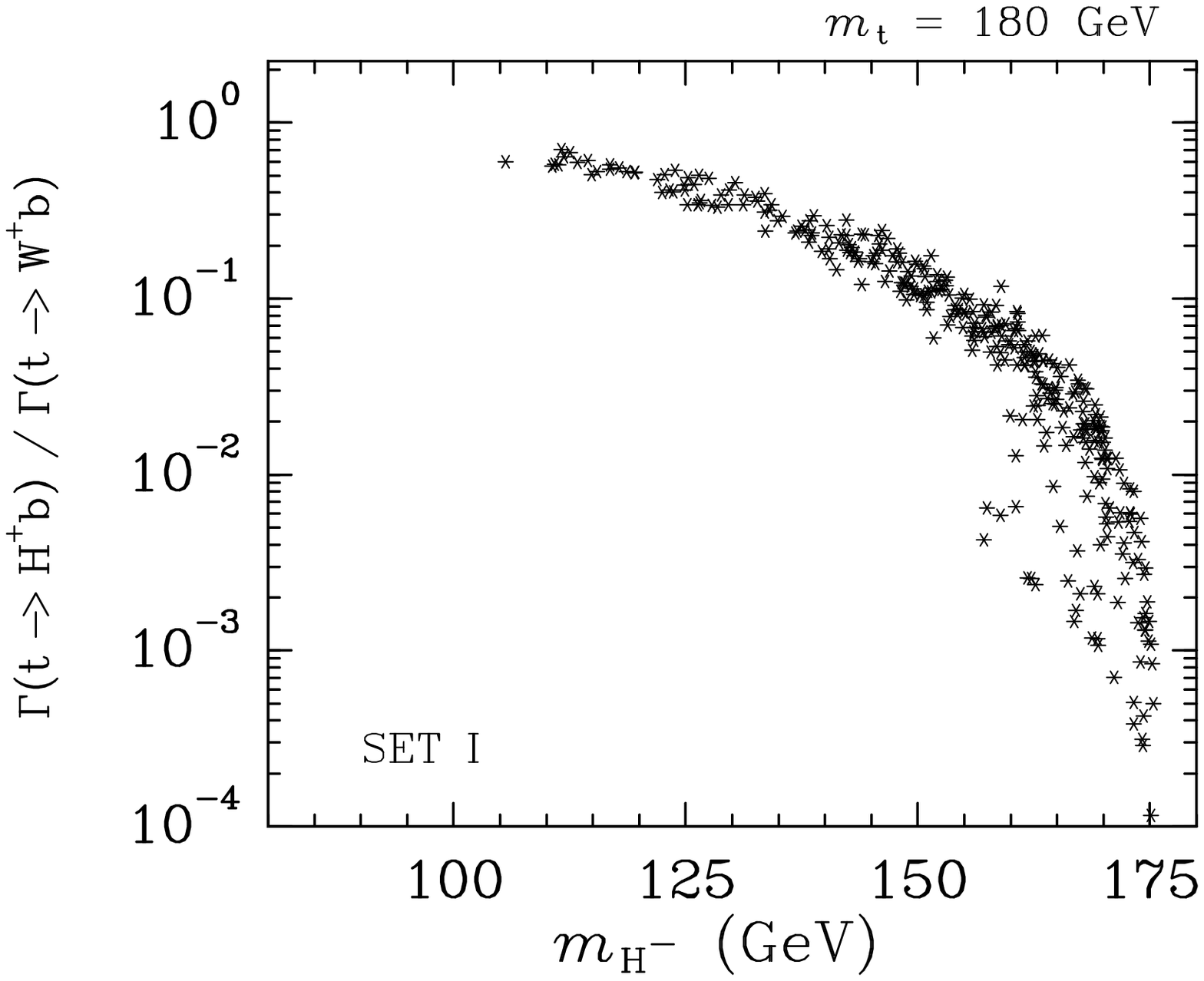}
\caption[f5]{\tenrm{Rates for $\thb$ in a generic 
model with SUSY type of Higgs-couplings and in the MSSM}}
\label{generic}
\end{figure}

\looseness=-1
Finally, we impose the constraints set by the CLEO Collaboration on 
$\bsg$: 
$1\cdot 10^{-4} \!<\! BR(\bsg) \!<\! 4.2 \cdot 10^{-4}$~\cite{CLEO}. 
The evaluation of $BR(\bsg)$ is in accordance to~\cite{BBMR}; the 
uncertainty associated to the theoretical prediction (due primarely 
to the ambiguity in the renormalization scale and to the experimental 
errors on parameters entering in the calculation) is estimated 
according to~\cite{BMMP}.

\vskip 0.5truecm 
{\elevenbf\noindent 2. Results: phase spaces and widths}
\vskip 0.2truecm 

\begin{figure}[h]
\epsfxsize=8.0 cm
\leavevmode
\epsfbox[60 360 550 745]{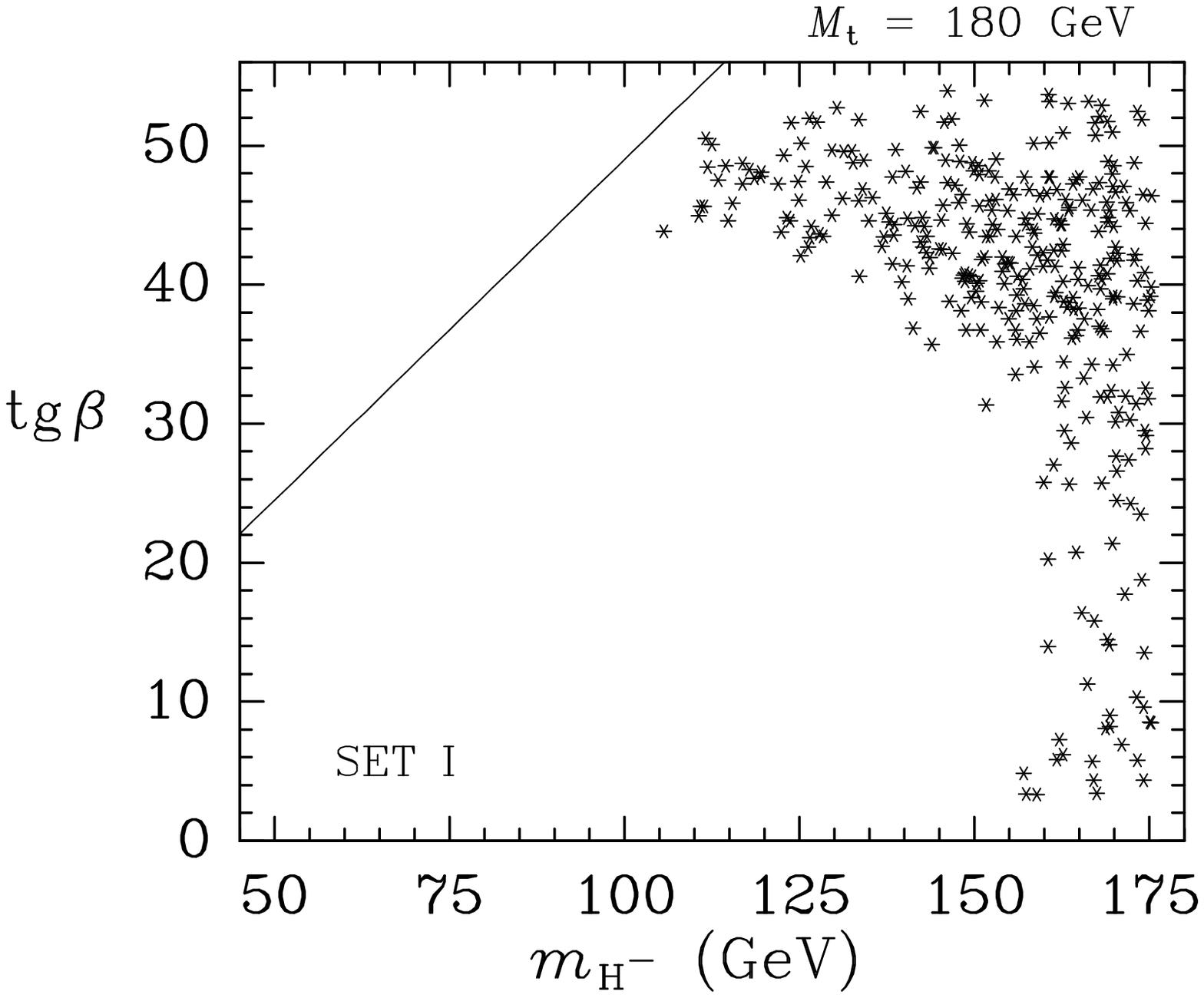}
\epsfxsize=8.0 cm
\epsfbox[30 360 530 745]{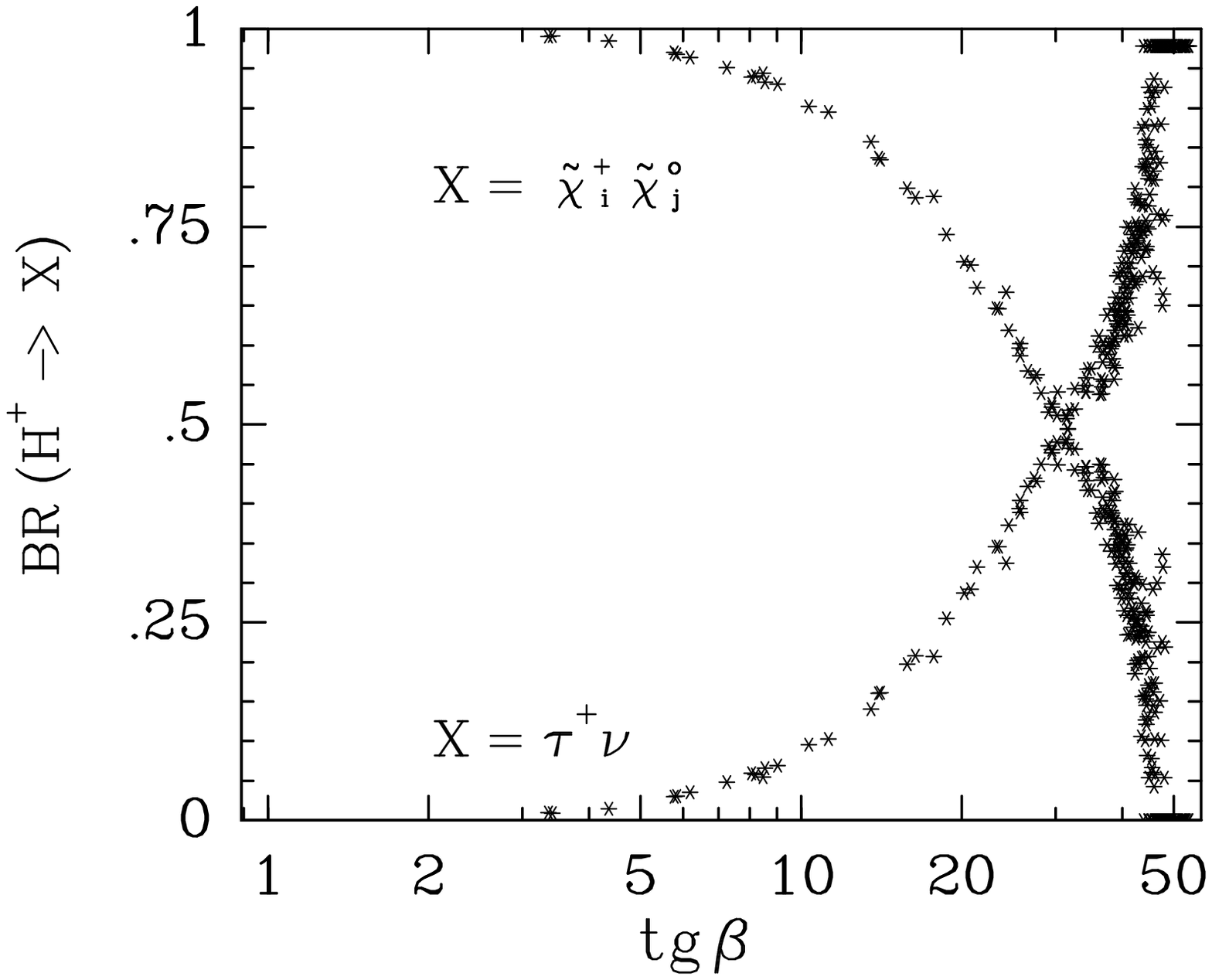}
\caption[f5]{\tenrm{Allowed phase space for $\thb$ and 
   branching ratios for the decays \Hcn\ , \Htn }}
\label{generic2}
\end{figure}
\looseness=-1
We present in the following the results which we obtain in the MSSM 
for the two decay modes $\thb$ and $\tstn$. No QCD or supersymmetric 
corrections are included in the calculation of the rates. We show our
results separetely before and after imposing the $\bsg$ constraints.

\looseness=-1
We show in the first frame of Fig.~\ref{generic} the $\mch$
and $\tanb$ dependence for the rate $\gamh /\gamw$ in a model with 
two Higgs doublets and type II Higgs couplings. The typical
minimum for intermediate values of $\tanb$ ($3\!<\!\tanb\!<\!9$) is
displayed, as well as the following steady increase for increasing
$\tanb$. The second frame shows the rates obtained in our simulation
when the experimental constraints SETI are imposed. The relative phase
space for this decay is shown in the first frame of
Fig.~\ref{generic2} where, for completness, we also display the line
$\tanb(\mbox{GeV})/\mch = 0.5$: the region excluded by the decay 
$b\to c \tau\nu_{\tau}$ lies above it. The $H^+$--masses shown in these
figures correspond, in general, to values of 
$\vert \mu\vert \sim 150\,$GeV and 
$150 \ltap \vert \mu \vert \ltap 300\,$GeV for $\tanb \ltap 30$.

\begin{figure}[h]
\vskip 0.4truecm 
\epsfxsize=8.0 cm
\leavevmode
\epsfbox[60 360 560 745]{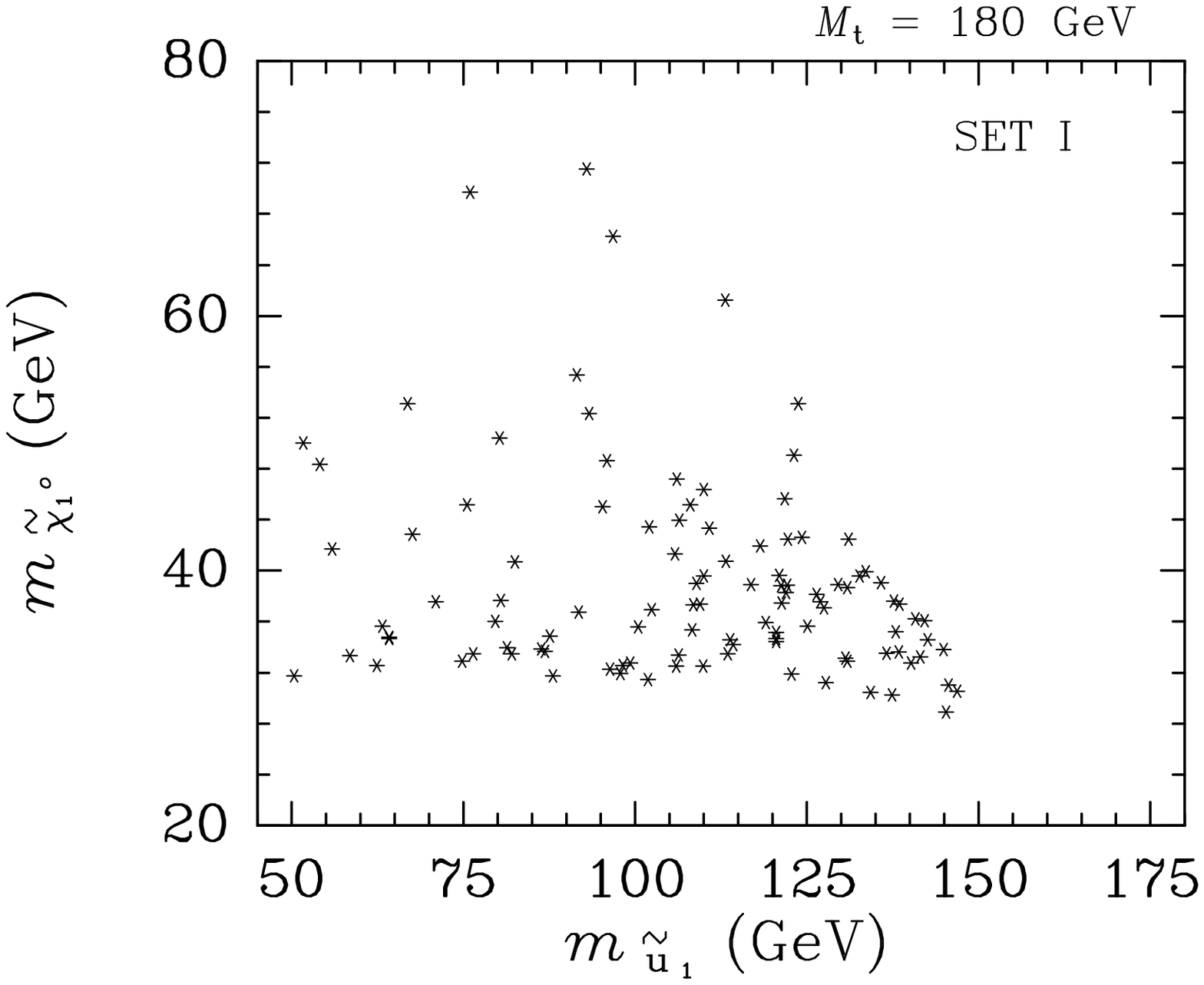}
\epsfxsize=8.0 cm
\epsfbox[30 360 530 745]{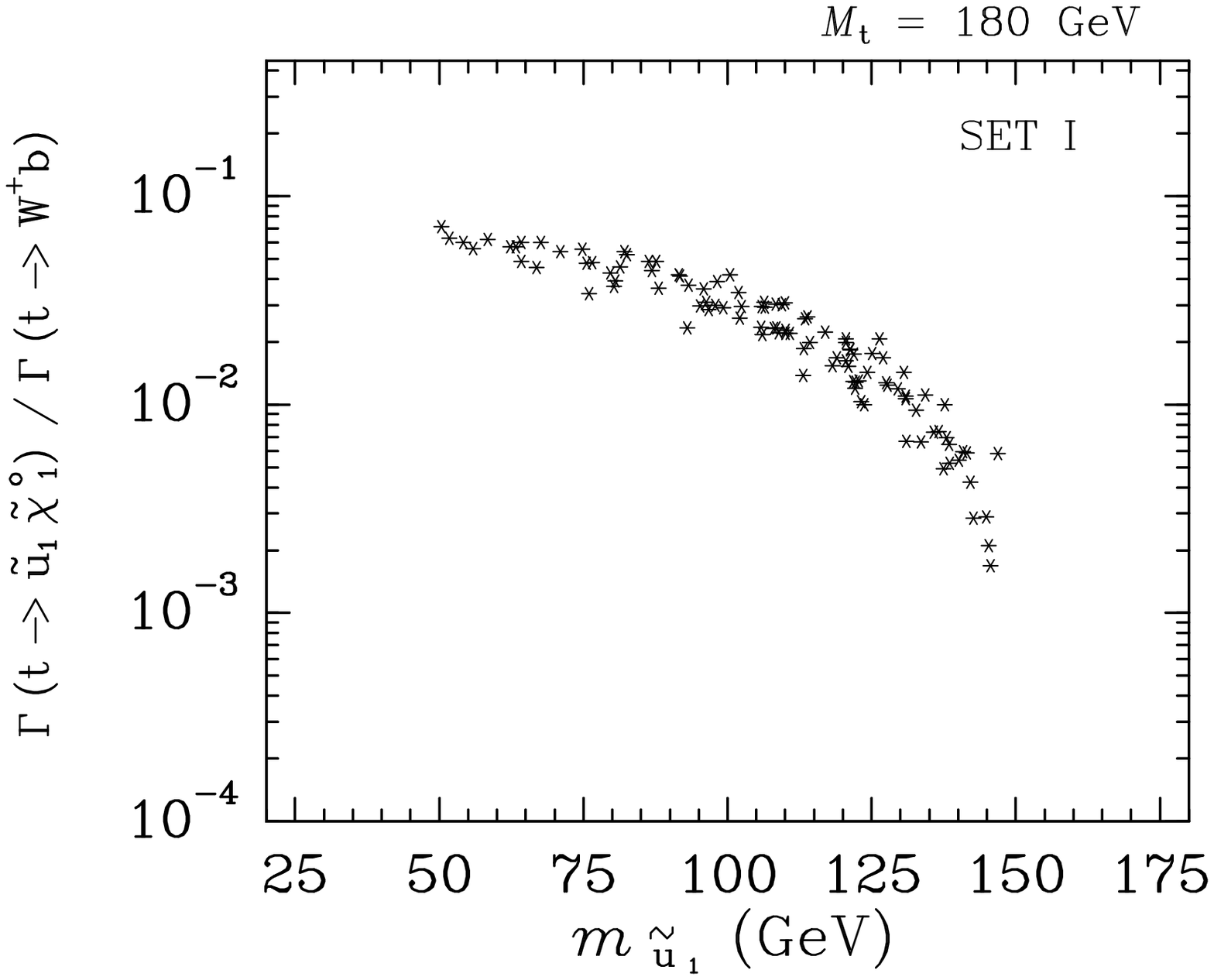}
\caption[f5]{\tenrm{Region of parameter space where 
$\tstn$ is kinematically accessible and relative rates}}
\label{tstn}
\end{figure}
\looseness=-1
The shape of the phase space obtained has not been drastically
affected by the inclusion of radiative corrections to the Higgs
potential, with respect to the phase space shown 
in~\cite{FBtop,FBbsg}. The reduction in the size of this area is 
due to the increase of the experimental lower bounds imposed in 
this search and to the slightly heavier spectrum of the 
Higgs--mass parameters used here. As in~\cite{FBtop}, we find 
that the largest rates are obtained for the largest values of 
$\tanb$ where also the lightest $H^+$ are found. The second frame 
of Fig.~\ref{generic2} shows the branching ratios for the two 
possible decay modes of the produced $H^+$. The mode 
$\tau \nu_\tau$ saturates the total width for $H^+$ only for 
very large \tanb. 

\looseness=-1
Wider regions of the supersymmetric parameter space need to be scanned
to obtain the points where the decay $\tstn$ is kinematically
accessible. They correspond to large A and large $\vert \mu \vert$ and
they are shown in Fig.~\ref{tstn}, together with the relative widths.
In the lower part of the region, the decay of ${\wti u}_1$ into 
an on-shell chargino is still possible. 
The set of viable points in this figure is disjoint from that where
$\thb$ is allowed, i.e. the corresponding values of $\mch$ in the
points of Fig.~\ref{tstn} are large, in general above
$250\,$GeV. Therefore, the non-observation of one of the two decay
modes, in general, does not affect the possibility of observing the
other one. 

\begin{figure}[t]
\epsfxsize=8.0 cm
\leavevmode
\epsfbox[60 360 560 745]{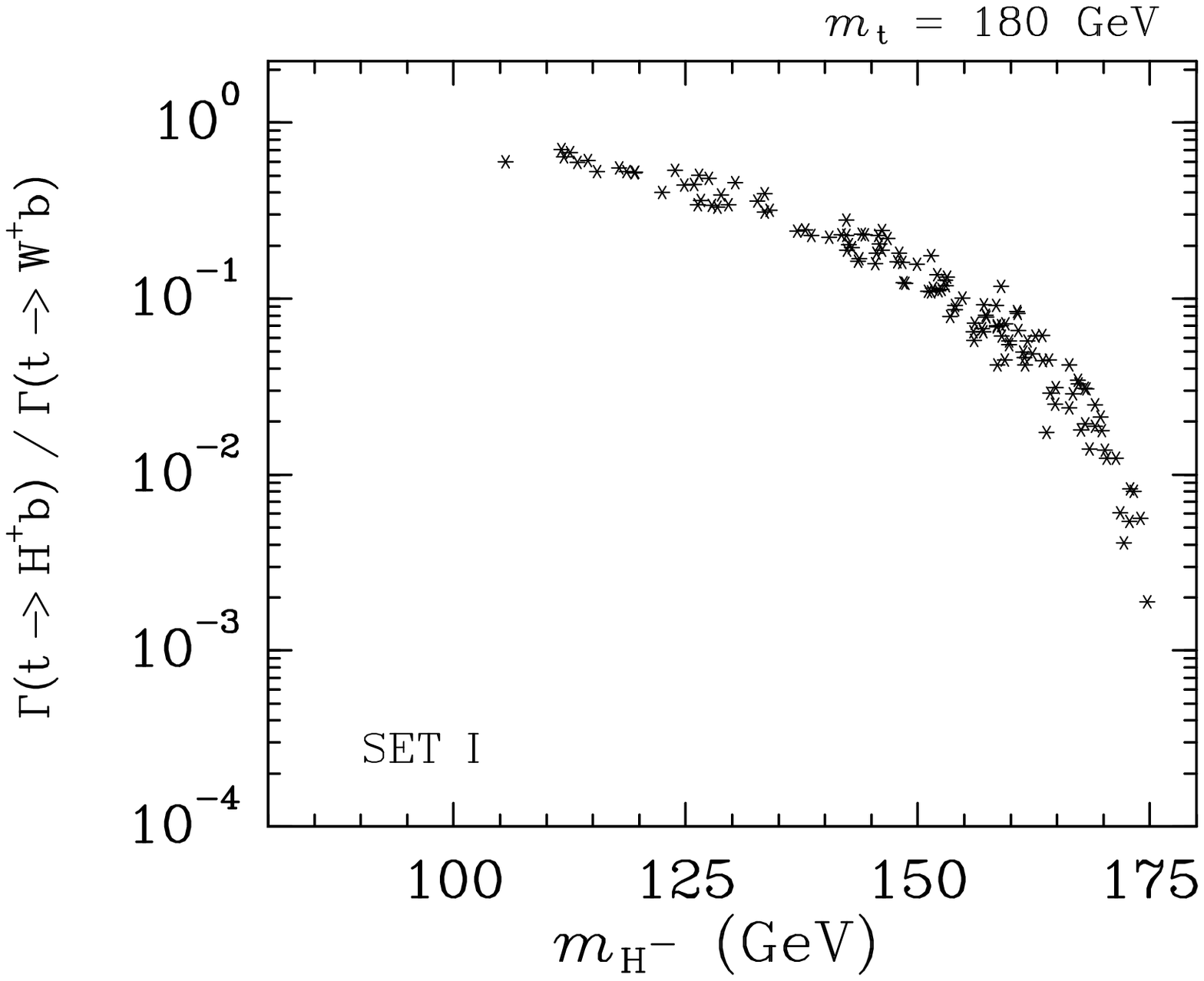}
\epsfxsize=8.0 cm
\epsfbox[30 360 530 745]{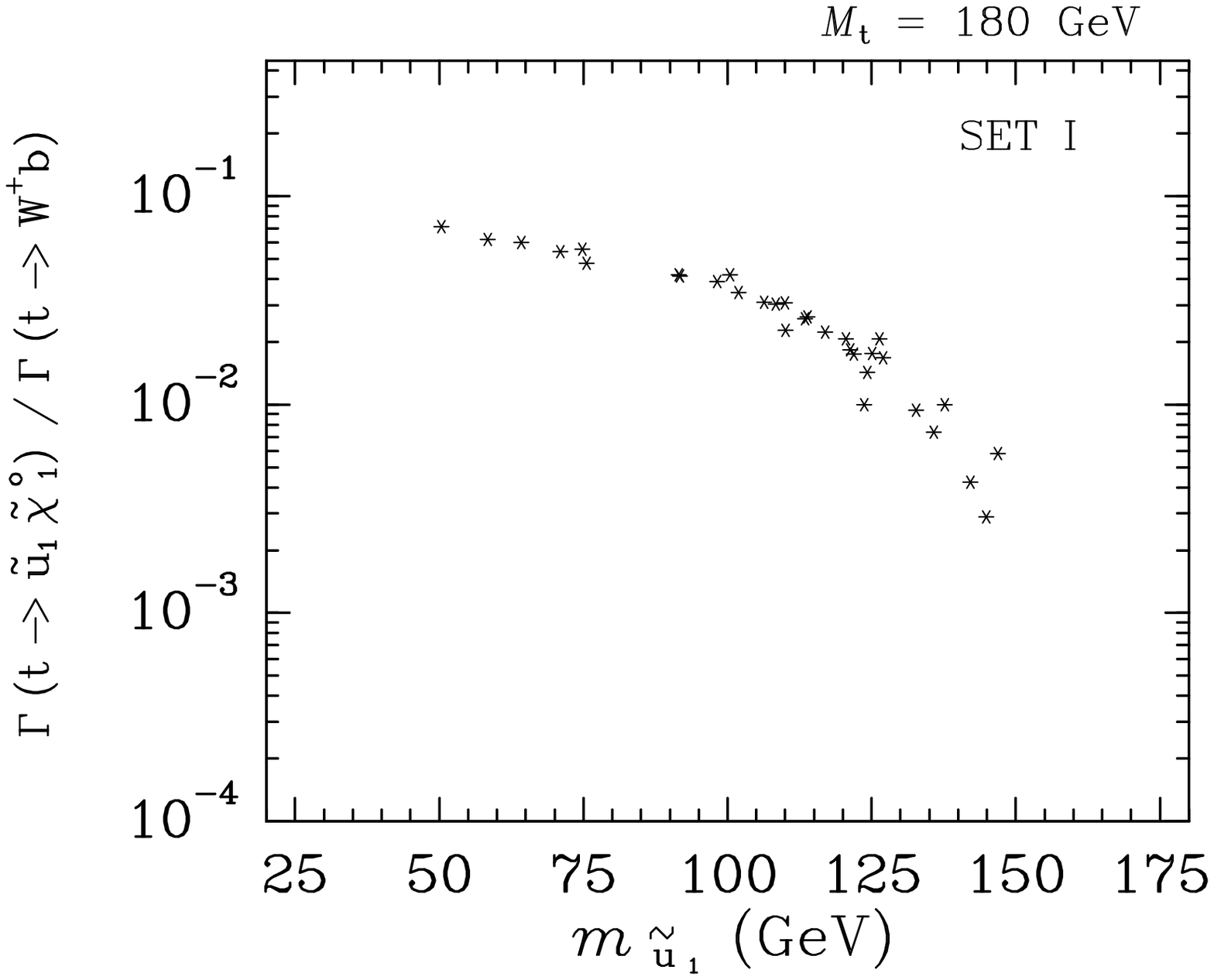}
\caption[f5]{\tenrm{Rates still reachable after the present
  experimental constraint on $\bsg$ is imposed}}
\label{bsg}
\end{figure}

\looseness=-1
We show in Fig.~\ref{bsg} what is left for the rates of the two decays
once we impose that the lower theoretical estimate of $BR(\bsg)$ is
$\!<\!4.2\cdot 10^{-4}$ and that the highest one is 
$\!>1\! \cdot 10^{-4}$, for each point of the MSSM parameter 
space (see for example discussion in~\cite{BDN}).
This figure summarizes our prediction for these rates in the MSSM when
all experimental constraints existing at present are imposed. We
should warn the reader, however, that of the two, the prediction for
$\thb$ is the most ``stable''. It does not change very much if the
band of allowed values for $\bsg$ is slightly restricted (see for 
example the band obtained at 95\% c.l. from the measurement
$BR(\bsg)= (2.32\pm0.57\pm0.35)\cdot 10^{-4}$~\cite{CLEO} 
when adding statistic and 
systematic errors in quadrature), whereas the allowed region for 
$\tstn$ tends to shrink.
 
\looseness=-1
Our results for $\thb$ are in qualitative agreement with those 
obtained in~\cite{OKADA} where, however, milder lower bounds on 
supersymmetric particles are imposed. 

\vskip 0.5truecm 
{\elevenbf\noindent 3. Discussion and Future Prospects}
\vskip 0.2truecm 

\begin{figure}[h]
\epsfxsize=8.0 cm
\leavevmode
\epsfbox[60 360 560 745]{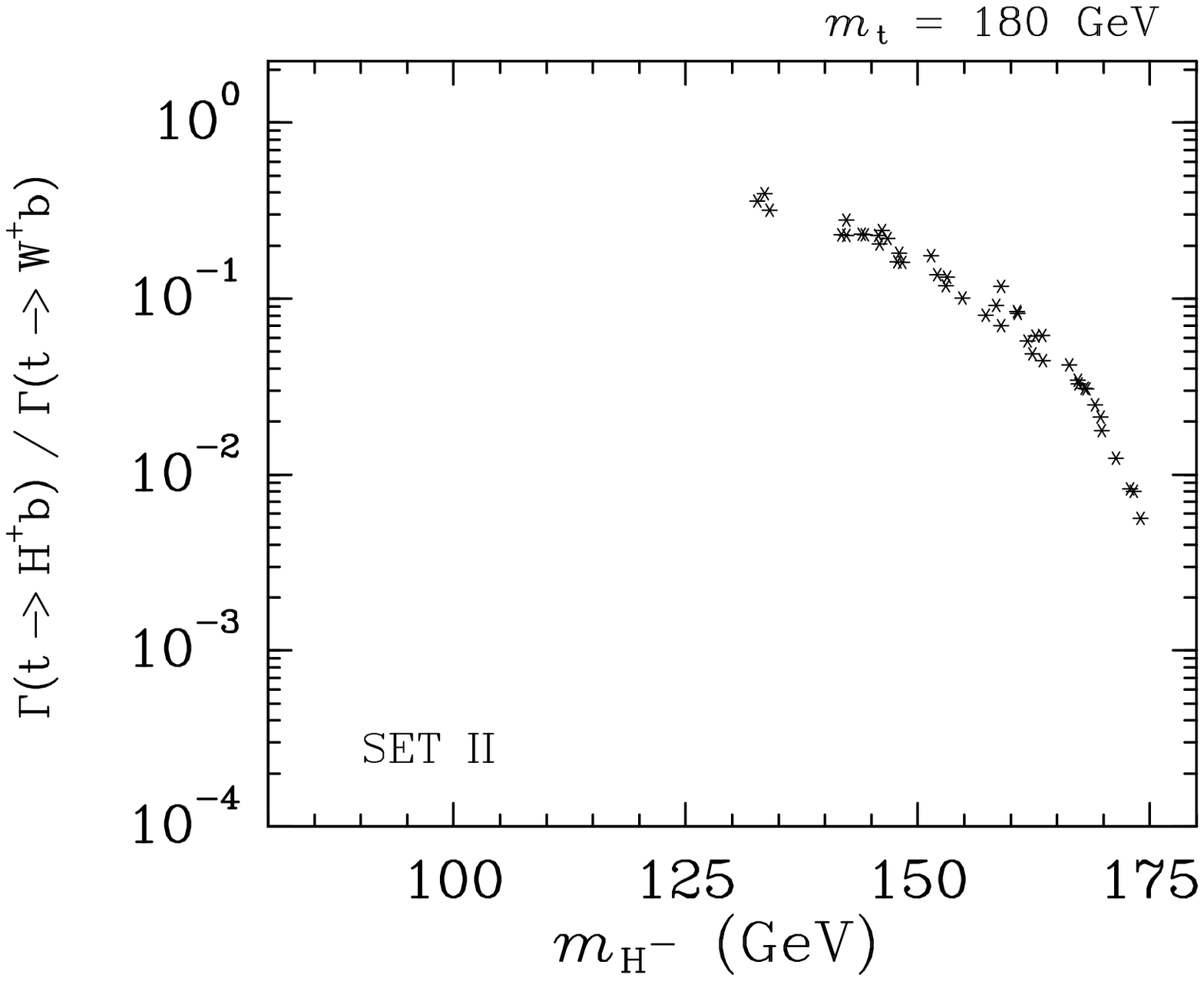}
\epsfxsize=8.0 cm
\epsfbox[30 360 530 745]{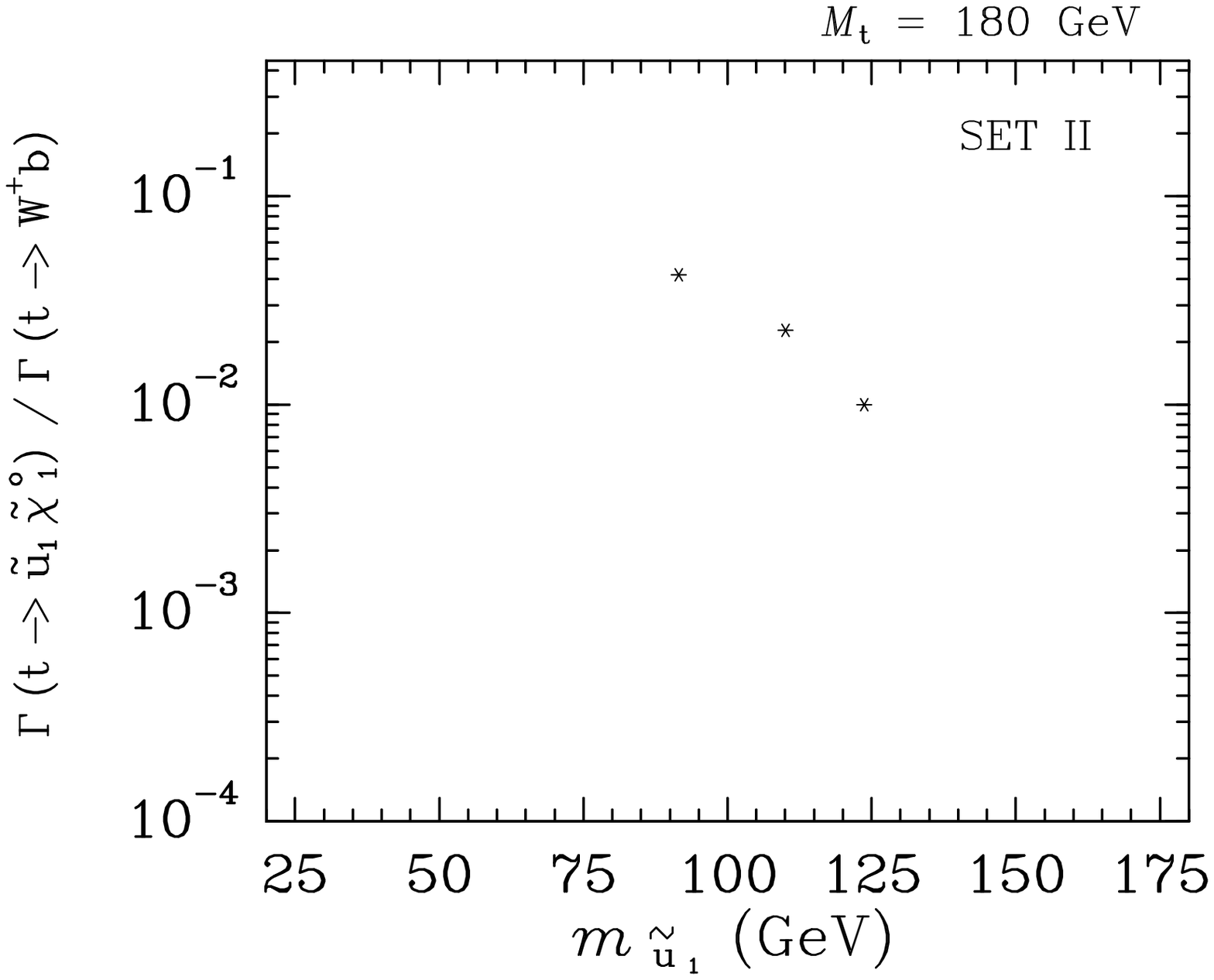}
\caption[f5]{\tenrm{Same as in Fig.~\ref{bsg} when the lower bounds 
  on supersymmetric masses SET~II are imposed}}
\label{future}
\end{figure}
\looseness=-1
As shown in the previous figures, rather large rates for $\thb$ can be
accomodated within the MSSM at the moment. The largest values for
$\Gamma(\thb)/\Gamma(\tstn)$ may turn out to be excluded 
in future searches at the TEVATRON.
More precision in the experimental measurement and in the theoretical
calculation of $BR(\bsg)$ (through the inclusion of NLO QCD
corrections) may completely close the second decay-mode $\tstn$. 
The same effect can 
be induced by an increase in the lower bounds on 
supersymmetric masses which LEP~II and
the TEVATRON will be able to put if no supersymmetric particle
are detected in the meantime. We show in Fig.~\ref{future} our
predictions for the two decays when the lower bounds SET~II are
imposed, together with the present limit on $\bsg$.

\looseness=-1
We conclude noticing that the consequences of the 
non-observation of $\thb$ will not be drastic for the MSSM (since 
the regions of parameter space where $\mch < M_t$ are indeed not very
large), whereas they may turn out to be interesting for models where
the strong correlations of the MSSM parameter space are absent.

\vskip 0.5truecm
{\elevenbf \noindent 4. Acknowledgements \hfill}

\noindent 
\looseness=-1
We acknowledge useful discussions with S.~Bertolini, M.~Carena,
A.~Djouadi, M.~Drees, Y.~Grossman, H.~Haber, R.~Hempfling, Y.~Nir, 
and F.~Vissani. We also thank G.~Wolf, L.~Lyons, E.~Kovacs, 
K.~Maeshima, and C.~Couyoumtzelis for providing informations 
concerning charged Higgs searches.

\end{document}